\documentclass[twocolumn,amsmath,amssymb,floatfix,prb,shownopacs,footinbib,superscriptaddress]{revtex4}
\usepackage{amsmath,amssymb,natbib,bm,graphicx,url,psfrag}
\usepackage[british]{babel}

\usepackage{color}

\newcommand{\bra}[1]{\left\langle  #1 \right\rvert}
\newcommand{\ket}[1]{\left\lvert #1  \right\rangle}

\newcommand{\com}[2]{\left[#1,#2\right]}

\newcommand{\DD}{\mathcal{D}}

\newcommand{\be}{\begin{equation}}
\newcommand{\ee}{\end{equation}}

\DeclareMathOperator{\tr}{tr}

\begin{document}

\title{Nonlinear response of the vacuum Rabi resonance}
\author{Lev S.\ Bishop}
\affiliation{Departments of Physics and Applied Physics, Yale University, PO Box 208120, New Haven, CT 06520, USA}
\author{J.\ M.\ Chow}
\affiliation{Departments of Physics and Applied Physics, Yale University, PO Box 208120, New Haven, CT 06520, USA}
\author{Jens Koch}
\affiliation{Departments of Physics and Applied Physics, Yale University, PO Box 208120, New Haven, CT 06520, USA}
\author{A.\ A.\ Houck}
\affiliation{Departments of Physics and Applied Physics, Yale University, PO Box 208120, New Haven, CT 06520, USA}
\author{M.\ H.\ Devoret}
\affiliation{Departments of Physics and Applied Physics, Yale University, PO Box 208120, New Haven, CT 06520, USA}
\author{E.\ Thuneberg}
\affiliation{Department of Physical Sciences, University of Oulu, PO Box 3000, FI-90014, Finland}
\author{S.\ M.\ Girvin}
\affiliation{Departments of Physics and Applied Physics, Yale University, PO Box 208120, New Haven, CT 06520, USA}
\author{R.\ J.\ Schoelkopf}
\email[email: ]{robert.schoelkopf@yale.edu}
\affiliation{Departments of Physics and Applied Physics, Yale University, PO Box 208120, New Haven, CT 06520, USA}

\date{\today}
\maketitle {\sffamily On the level of single atoms and photons, the coupling between atoms and the
electromagnetic field is typically very weak. By employing a cavity to confine the field, the strength of this
interaction can be increased many orders of magnitude to a point where it dominates over any dissipative
process. This strong-coupling regime of cavity quantum electrodynamics (QED)\cite{raimond_colloquium:_2001,kimble}, has
been reached for real atoms in optical cavities\cite{thompson_observation_1992}, and for artificial atoms in circuit QED\cite{wallra_strong_2004} and quantum-dot systems\cite{reithmaier_strong_2004,yoshie_vacuum_2004}.
A signature of strong coupling is the splitting of the cavity transmission peak into a pair of resolvable peaks
when a single resonant atom is placed inside the cavity -- an effect known as vacuum Rabi splitting. The
circuit QED architecture is ideally suited for going beyond this linear response effect. Here, we show that
increasing the drive power results in two unique nonlinear features in the transmitted heterodyne signal: the
supersplitting of each vacuum Rabi peak into a doublet, and the appearance of additional peaks with the
characteristic $\sqrt{n}$ spacing of the Jaynes--Cummings ladder. These constitute direct evidence for the
coupling between the quantized microwave field and the anharmonic spectrum of a superconducting qubit acting as
an artificial atom.}

Circuit QED realizes the coupling between a superconducting qubit  and the microwave field inside an on-chip
transmission line resonator\cite{blais_cavity_2004,wallra_strong_2004}. The drastic reduction in mode volume for
such a quasi-1d system \cite{devoret_circuit-qed:_2007,schoelkopf_wiring_2008}, the recent improvement in
coherence times\cite{schreier_suppressing_2008}, and the absence of atomic motion render circuit QED an ideal
system for studying the strong-coupling limit. In the customary linear-response measurement of the transmitted
microwave radiation, the vacuum Rabi splitting manifests itself as an avoided crossing between qubit and
resonator. 

It may be argued that the observation of the splitting is not yet a clear sign for quantum behaviour
\cite{zhu_vacuum_1990,tian_quantum_1992}, since avoided crossings are also ubiquitous in classical physics.
However, quantum mechanics gives rise to a distinct nonlinearity of these splittings when initializing the
resonator in a higher photon Fock state: when the resonator mode is occupied by $n$ photons, the splitting is
enhanced by a factor $\sqrt{n+1}$ as compared to the vacuum Rabi situation. This nonlinearity has  been observed
with single atoms in both microwave\cite{brune_quantum_1996} and optical
cavities\cite{birnbaum_photon_2005,schuster_nonlinear_2008}. In circuit QED, this characteristic trait has been
observed very recently both spectroscopically\cite{wallraff}, as well as in time-domain measurements\cite{simmonds,martinis}.

Here, we present a theoretical analysis and experimental investigation of the vacuum Rabi resonance in a circuit
QED system, where we study the $\sqrt{n}$ nonlinearity up to $n=5$ by exploring the power dependence of the
heterodyne transmission (see Methods). This type of detection is particularly simple,
since it is a continuous measurement involving only a single driving frequency. We find that, surprisingly, each
vacuum Rabi peak develops into a  new doublet (supersplitting), which can be simply explained by the saturation
of a two-level system comprising both photon and qubit degrees of freedom. For even stronger driving,
a series of additional peaks due to multiphoton transitions\cite{solano} up the Jaynes--Cummings ladder appear, but the complete shape of
the spectrum closely matches the predictions from  a master equation treatment including only relaxation, dephasing, and
the known multilevel spectrum of the superconducting qubit. The quantitative agreement confirms the validity of
the extended Jaynes--Cummings Hamiltonian  for our system, rendering circuit QED an excellent tool for probing
the fundamental interaction of matter and light.

\begin{figure}
 \centering
 \includegraphics[width=1.0\columnwidth]{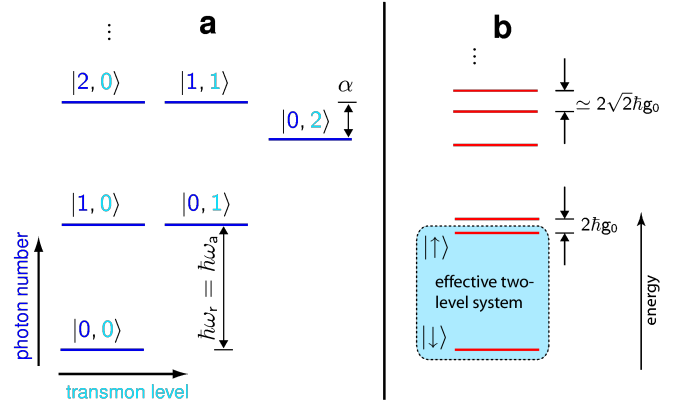}
 \caption{{\bfseries Extended Jaynes--Cummings level diagram of the resonator--transmon system}. {\bfseries a},~Bare
 levels in the absence of coupling. The states are denoted $\ket{n,j}$ for photon
 number $n$ and occupation of transmon level $j$.
 {\bfseries b},~Spectrum of the system including the effects of transmon--resonator coupling. The effective two-level system
 relevant to describing the lower vacuum Rabi peak comprises
 the ground state and the antisymmetric combination of qubit and photon excitations.
 \label{fig:fig1}}
 \end{figure}
We consider a system consisting of a superconducting transmon
qubit\cite{koch_charge-insensitive_2007,schreier_suppressing_2008} whose multiple quantum levels are coupled to
a single mode of a transmission-line resonator. Our arguments, however, will be general and should apply to any
other realization of cavity QED with sufficiently large coupling and atom coherence times. Taking into account
the relatively small anharmonicity of the transmon qubit, the system is modelled in terms of an extended
Jaynes--Cummings hamiltonian including a coherent drive at frequency $\omega_\text{d}/2\pi$,
\begin{align}
H&=\hbar\omega_\text{r} a^\dag a +  \sum_j \hbar\omega_j \ket{j}\bra{j}
\\\nonumber
&+\sum_{j}\hbar g_j\Bigl (a \ket{j+1}\bra{j}+\text{h.c.} \Bigr)+\hbar\xi (a^\dag e^{-i\omega_\text{d}t}+\text{h.c}).
\end{align}
Here, the operator $a$ ($a^\dag$) annihilates (creates) a photon in the resonator mode with frequency
$\omega_\text{r}/2\pi$. The energies corresponding to the transmon eigenstates are $\hbar \omega_j$
($j=0,1,\ldots$), and they are determined by the Cooper pair box hamiltonian\cite{bouchiat_quantum_1998}  for
given Josephson energy $E_J$ and charging energy $E_C$. In the transmon regime considered here, $E_J/E_C\gg1$,
the levels form a weakly anharmonic ladder with anharmonicity $\alpha=\omega_{12}-\omega_{01}\sim -E_C$
($\omega_{ij}=\omega_j-\omega_i$ is the transition frequency between levels $i$ and $j$). The
transmon--resonator coupling strengths are  $g_j\sim g_0\sqrt{j+1}$, where $g_0/2\pi$ is the vacuum Rabi
frequency. The resulting energy level diagram for vanishing drive, $\xi=0$, is schematically shown in
Fig.~\ref{fig:fig1}.

The interaction of the cavity--transmon system with its environment enables control and measurement by means of microwave drive and detection\cite{blais_cavity_2004,wallra_strong_2004}, but also leads to decoherence. The appropriate framework to capture all these effects is a description in terms of the reduced density matrix $\rho$ for the cavity and transmon. The Markovian master equation which governs the evolution of $\rho$ is given by
\begin{align}\label{master2}
\dot\rho=&-\frac{i}{\hbar}\com{H}\rho+\kappa\DD[a]\rho
            +\gamma_{1}\DD\Bigl[\sum_{j}\alpha_{j}\ket{j}\bra{j+1}\Bigr]\rho\nonumber\\
            &+\frac{\gamma_\varphi}{2} \DD\Bigl[\sum_j \beta_j\ket{j}\bra{j}\Bigr]\rho ,
\end{align}
where $\DD[A]$ is the usual Lindblad damping superoperator defined by
$\DD[A]\rho=\left(\com{A\rho}{A^\dag}+\com{A}{\rho
A^\dag}\right)/2$. The three damping terms on the right-hand side model the loss of cavity photons at rate $\kappa$, the intrinsic relaxation of transmon excitations at rate $\gamma_{1}$, and the pure dephasing of transmon state superpositions at rate $\gamma_\varphi$ and relative strengths $\alpha_j$, $\beta_j$ (see Methods).

\begin{figure}
 \centering
 \includegraphics[width=1.0\columnwidth]{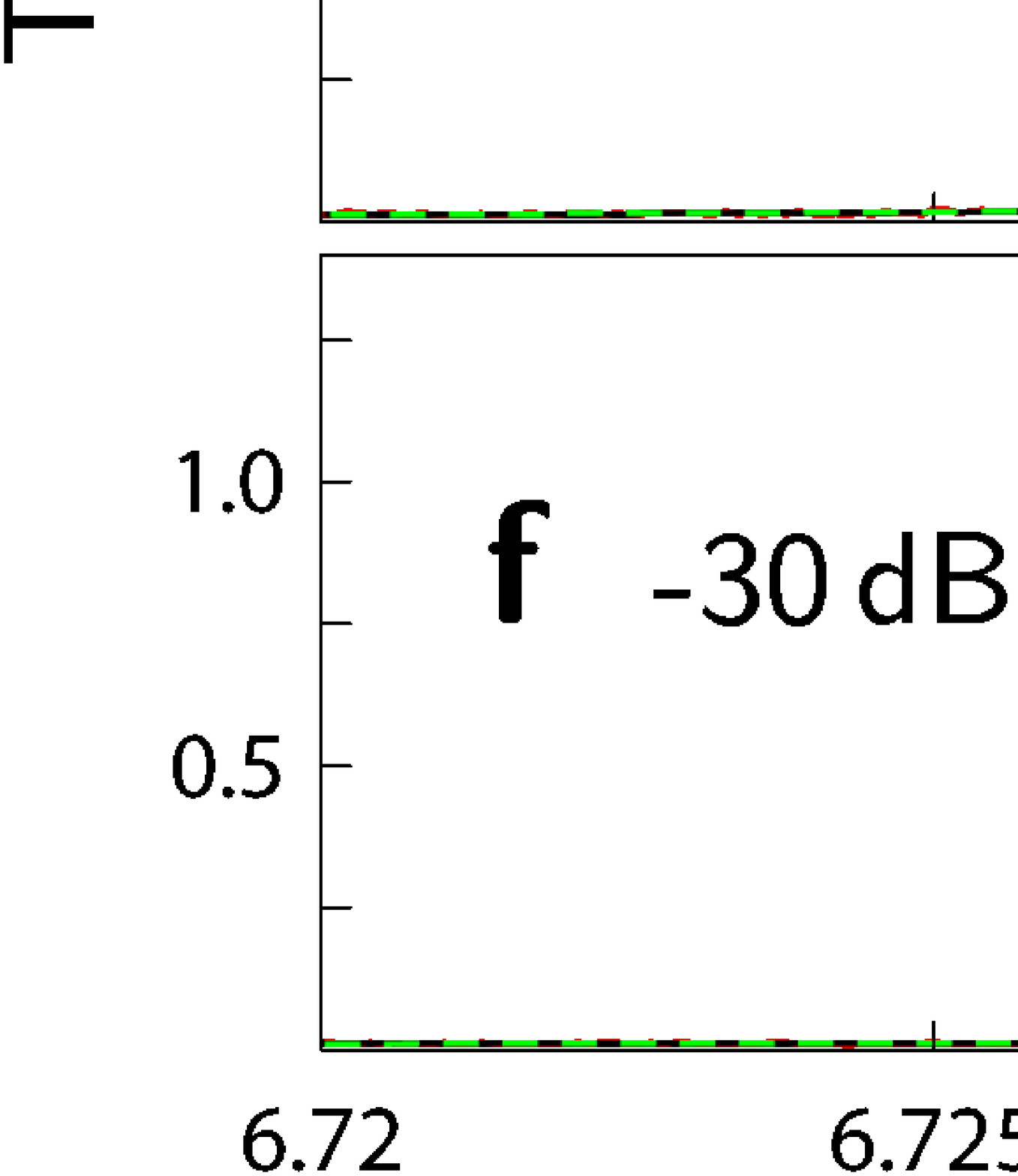}
 \caption{{\bfseries Supersplitting of the vacuum Rabi resonance when probing heterodyne transmission beyond linear response.} The experimental data are obtained with a circuit QED system in the strong-coupling regime, where the vacuum Rabi splitting is observed to exceed 260 linewidths, see {\bfseries a}.  All plots show the
 squared heterodyne amplitude $A^2$ in arbitrary units.
 {\bfseries b},~Measured transmission (colour scale) for the left vacuum Rabi peak, as a function of drive frequency and power.
  The plot reveals the supersplitting of a single Lorentzian into a doublet of peaks.
 {\bfseries c--f},~Cuts for constant power at the values indicated in {\bfseries b}. In linear response, {\bfseries f}, the vacuum Rabi peak is Lorentzian; as the power increases a central dip
 develops, {\bfseries e}, leading to supersplitting of the peak, {\bfseries d}, and eventually becoming asymmetric at the largest powers, {\bfseries c}. The experimental data (red line) is in excellent agreement with theory (black line). The results from the 2-level approximation are shown for comparison (green dashed line).
 \label{fig:exp}}
 \end{figure}

The quantity of interest for detection of the vacuum Rabi splitting is the transmission amplitude $A(\omega_\text{d})$ of a microwave signal at frequency $\omega_\text{d}/2\pi$. Here, we employ a heterodyne measurement of the field quadratures $I=\langle b_\text{out}+b_\text{out}^\dagger\rangle$ and $Q=\langle ib_\text{out}^\dagger-ib_\text{out}\rangle$, where $b_\text{out}$ refers to the output field at the drive frequency. Input-output theory\cite{walls_quantum_1995} enables us to express the steady-state transmission amplitude as
\begin{equation}\label{ampl}
A=\sqrt{I^2+Q^2}=2\sqrt{\kappa}\lvert\langle a\rangle\rvert = 2\sqrt{\kappa}\lvert \tr (a\,\rho_\text{s}) \lvert.
\end{equation}
Our calculations thus only require the steady-state solution $\rho_\text{s}$ of the master equation
\eqref{master2}. We explore the transmission intensity $A^2$ as a function of the drive strength $\xi$, and
compare our results to experimental data obtained for a transmon qubit in resonance with a
$\omega_\text{r}/2\pi=6.92\,\text{GHz}$ coplanar waveguide (CPW) resonator with coupling strength
$g_0/\pi=347\,\text{MHz}$ (see Methods for experimental details). In the linear-response regime, the vacuum Rabi
peaks have the characteristic Lorentzian line shape. Their separation and width are given by $2g_0$ and
$(\gamma_1+2\gamma_\varphi+\kappa)/2$, respectively. For our specific sample, the splitting is observed to exceed
260 linewidths, shown in Fig.\ \ref{fig:exp}a. 
Strikingly, when increasing the drive power beyond the linear-response regime, the shape of the transmission curve changes drastically as shown in Figs.\ \ref{fig:exp}b--f. Each vacuum Rabi peak develops a central dip and eventually supersplits into a doublet of peaks. We find excellent agreement between our prediction based on Eqs.\ \eqref{master2} and \eqref{ampl} and the measured transmission.

The physics of this supersplitting can be elucidated within a reduced two-level model. This model takes into
account only the Jaynes--Cummings ground state and either the symmetric or antisymmetric superposition of
transmon and photon excitation. These two states, $\ket{\downarrow}=\ket{0,0}$ and
$\ket{\uparrow}=(\ket{1,0}\pm\ket{0,1})/\sqrt{2}$, form an effective two-level system composed of both transmon
and cavity degrees of freedom. Within the effective two-level subspace, the photon operators are mapped\cite{carmichael_statistical_2008}
to Pauli operators $a\to\Sigma_-/\sqrt{2}$, $a^\dag\to\Sigma_+/\sqrt{2}$, so that the
microwave tone acts as a drive on the effective two-level system, i.e.\
\begin{equation}\label{Htilde}
H_\text{eff} = \frac{\hbar\Delta}{2}\Sigma_z+\frac{\hbar\Omega}{2}\Sigma_x,
\end{equation}
a scenario that Carmichael et al.\cite{tian_quantum_1992,carmichael_statistical_2008} have referred to as ``dressing of dressed states."
The hamiltonian $H_\text{eff}$ refers to the frame rotating at the drive frequency, $\Delta=\omega_{01}\pm g_0-\omega_\text{d}$ is the detuning between drive and one vacuum Rabi peak, and $\Omega=\sqrt{2}\xi$ is the effective drive strength. With the notable exception of the recent work by I.\ Schuster et al.\cite{schuster_nonlinear_2008}, previous investigations were primarily concerned with effects on photon correlations and fluorescence, as observed in photon-counting measurements \cite{tian_quantum_1992,birnbaum_photon_2005}. According to the operator mapping, photon counting can be related to the measurement of $\langle \Sigma_z \rangle$, whereas detection of the heterodyne transmission amplitude $A$ corresponds to $|\langle \Sigma_-\rangle |$. As a result, heterodyne detection fundamentally differs from photon counting and the vacuum Rabi supersplitting is a characteristic of heterodyne detection only.

After restricting the master equation \eqref{master2} to the two-level subspace, the system evolution can be
expressed in terms of simple Bloch equations for the three components of the reduced density matrix
$\rho=(\openone+x\Sigma_x+y\Sigma_y+z\Sigma_z)/2$,
\begin{align}
\dot{x}&=-x/T'_2-\Delta y ,\qquad
\dot{y}=\Delta x-y/T'_2-\Omega z ,\nonumber\\
\dot{z}&=\Omega y -(z+1)/T'_1 .
\end{align}
(An intuitive approach avoiding the Bloch equations is discussed in the Supplementary Information, Discussion.)
Here, $T'_1$ and $T'_2$ are the effective relaxation and dephasing times, which are related to $\gamma_1$,
$\gamma_\varphi$, and $\kappa$ via ${T'_1}^{-1}=(\gamma_1+\kappa)/2$, and
${T'_2}^{-1}=(\gamma_1+2\gamma_\varphi+\kappa)/4$. From the steady-state solution of the Bloch equations we find
the transmission amplitude
\begin{equation}
\label{blochsol} A = \frac{T'_2\Omega\sqrt{\kappa(\Delta^2{T'_2}^2+1)/2}}
   {\Delta^2{T'_2}^2 + T'_1 T'_2 \Omega^2+1} .
\end{equation}
This expression correctly describes the crossover from linear response at small driving strength, $ \Omega\ll
(T'_1 T'_2)^{-1/2}$, producing a Lorentzian of width $2 {T'_2}^{-1}$, to the doublet structure observed for
strong driving. Specifically, as the drive power is increased, the response saturates and the peak broadens,
until at $\Omega=(T'_1 T'_2)^{-1/2}$ the peak undergoes supersplitting with peak--peak separation
$2{T'_2}^{-1}\sqrt{T'_1 T'_2 \Omega^2-1}$. The fact that we use heterodyne detection is indeed crucial for the
supersplitting. It is easy to verify within the two-level approximation that photon-counting always results in a
Lorentzian. For photon counting, probing beyond the linear-response regime merely results in additional
power-broadening; specifically, the width of the Lorentzian is given by $2{T'_2}^{-1}\sqrt{T'_1T'_2\Omega^2+1}$.

 In Figs.\ \ref{fig:exp}c--e, the analytical expression \eqref{blochsol} is plotted for comparison with the full numerical results and the experimental data. We find good agreement for low to
moderate drive power, confirming that the supersplitting can be attributed to driving the vacuum Rabi transition
into saturation while measuring the transmission with the heterodyne technique.  For higher drive power a
left-right asymmetry appears in the true transmission spectrum, which is not reproduced by Eq.~\eqref{blochsol},
and which is partly due to the influence of levels beyond the two-level approximation.

\begin{figure}
 \centering
 \includegraphics[width=1.0\columnwidth]{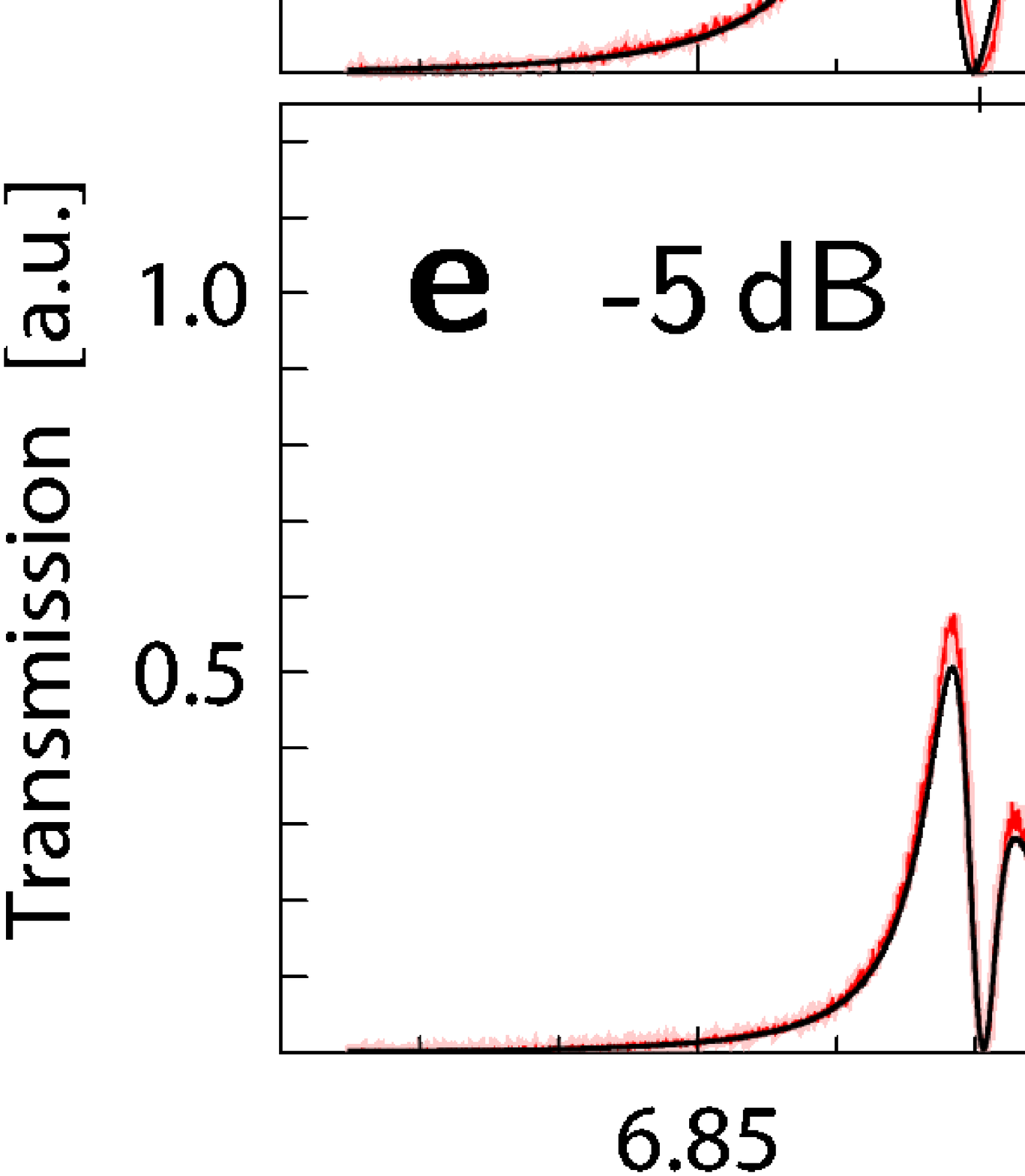}
\caption{{\bfseries Emergence of $\sqrt{n}$ peaks under strong driving of the vacuum Rabi transition.} {\bfseries a},~The extended
Jaynes--Cummings energy spectrum. All levels are shown to scale in the left part of the diagram: black lines
represent levels $\ket{n,\pm}\simeq(\ket{n+1,0}\pm\ket{n,1})/\sqrt{2}$ with only small contributions from higher
($j>1$) transmon states; grey lines represent levels with large contributions from higher transmon states. In
the right part of the diagram, the $\sqrt{n}$ scaling of the splitting between the $\ket{n,\pm}$ states is
exaggerated for clarity, and the transitions observed in plots {\bfseries b--e} are indicated at the $x$-coordinate
$E_{n\pm}/n\hbar$ of their $n$-photon transition frequency from the ground state. {\bfseries b},~Measured transmission
($A^2$, heterodyne amplitude squared) in colour scale as a function of drive frequency and power. The
multiphoton transitions shown in {\bfseries a} are observed at their calculated positions. {\bfseries c--e},~Examples of cuts for
constant power, at the values indicated in {\bfseries b} (results from the master equation~\eqref{master2} in black;
experimental results in red), demonstrating excellent agreement between theory and experiment, which is
reinforced in the enlarged insets. Good agreement is found over the full range in drive power from
$-45\,\text{dB}$ to $+3\,\text{dB}$, for a single set of parameters (also see Supplementary Movie~1).
\label{fig:powsweep}}
\end{figure}
Higher levels of the extended Jaynes--Cummings hamiltonian become increasingly important when the drive power is
raised further. Figure\ \ref{fig:powsweep} shows the emergence of additional peaks in the transmission spectrum.
Each of the peaks can be uniquely identified with a multiphoton transition from the ground state to an excited
Jaynes--Cummings state. For simplicity, we consider the situation where the anharmonicity $\alpha$ and the
coupling strength $g_0$ are sufficiently different that mixing between higher transmon levels and the regular
Jaynes--Cummings states $\ket{n,\pm}=(\ket{n+1,0}\pm\ket{n,1})/\sqrt{2}$ is minimal for the low-excitation
subspaces. Accordingly, the experiments are carried out using a different transmon qubit in the same sample,
with a smaller coupling of $g_0/\pi=94.4\,\text{MHz}$ (see Methods). In this case, the $n$-photon transitions to
the $n$-excitation subspace occur at frequencies $E_{n\pm}/nh=(\omega_\text{r}\pm n^{-1/2}g_0)/2\pi$, and thus
reveal the nonlinearity of the Jaynes--Cummings ladder. The detailed comparison between experimental data and
numerical simulation in Fig.\ \ref{fig:powsweep}c--e shows superb agreement down to the narrowest features
observed.

\begin{figure}
 \centering
 \includegraphics[width=1.0\columnwidth]{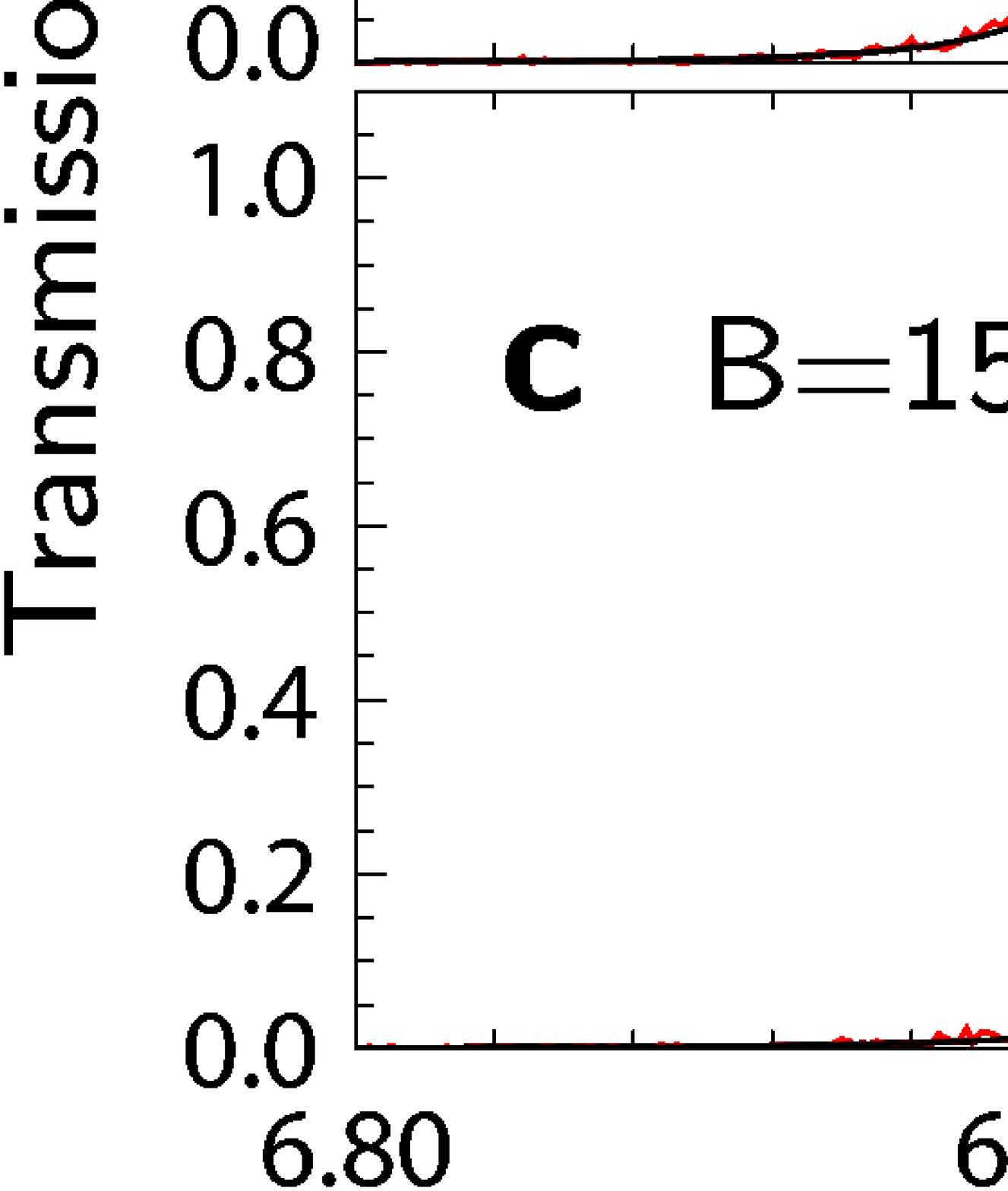}
\caption{{\bfseries Qubit--cavity avoided crossing at high drive power.} Transmission measurement when tuning the qubit frequency through resonance for a drive power of
$+1\,\text{dB}$. {\bfseries a},~Measured transmission as a function of drive frequency and magnetic field. As the field is
increased, the qubit frequency is tuned through resonance with the cavity, and anticrossing behaviour is
observed. The multiphoton transitions shown in Fig.~\ref{fig:powsweep}a are visible. The anomaly at $B\simeq
15.59$ is most likely due to the crossing of a higher level of the second qubit present in the same cavity.
{\bfseries b--c},~Example cuts at constant magnetic field, at the values indicated in {\bfseries a} (master equation results,
calculated using the same parameters as for Fig.~\ref{fig:powsweep}, are shown in black; measured results in
red). (Also see Supplementary Movie~2).
 \label{fig:wolverine}}
 \end{figure}

The possibility of multiphoton transitions at sufficiently large drive powers also affects the shape of the
vacuum Rabi splitting when tuning the qubit frequency $\omega_{01}$ through resonance with the cavity, shown in
Fig.~\ref{fig:wolverine}. Instead of the simple avoided crossing commonly observed at low drive
powers\cite{wallra_strong_2004}, the presence of multiphoton transitions leads to a fan-like structure where
individual branches can again be identified one-to-one with the possible transitions in the Jaynes--Cummings
ladder. In the experimental data of Fig.\ \ref{fig:wolverine}a, processes up to the 5-photon transition are
clearly visible. Detailed agreement with the theory verifies that the more general situation of non-zero
detuning between qubit and resonator is correctly described by our model.

In summary, we have shown the supersplitting and the emergence of multiphoton transitions in the nonlinear
response of the vacuum Rabi resonance, presenting both theoretical predictions and their experimental
verification.  This enables the observation of the characteristic $\sqrt{n}$ nonlinearities in the
Jaynes--Cummings ladder in a heterodyne transmission measurement, and is direct evidence for the strong coupling
between the superconducting qubit and the quantized microwave field. The high  precision of agreement between
predictions and experiment demonstrates the robustness of the Jaynes--Cummings physics in a circuit QED system.
This opens up pathways for further investigations in quantum optics and quantum information, such as generating
number-squeezed states or employing multi-level quantum logic.

\subsection*{METHODS}
{\noindent\sffamily THEORY}\\
The modelling of the transmon follows ref.\ \onlinecite{koch_charge-insensitive_2007}. While some expressions
given in the main text are asymptotic results valid for $E_J/E_C\gg1$, our calculations are based on a full
diagonalization of the Cooper pair box Hamiltonian.

The description of the transmon--cavity system in terms of a master equation requires a model for the relaxation
and dephasing of higher transmon levels. As detailed studies of the microscopic origin of the dominant
relaxation and dephasing channels are still outstanding, we have chosen plausible superoperators for our master
equation \eqref{master2}. Assuming that relaxation of higher transmon levels may arise due to a coupling of
external degrees of freedom to the charge on the superconducting island, we take the relative strengths of
relaxation to be related to the coupling parameters as $\alpha_j=g_j/g_0$. Dephasing of higher levels is likely
to be due to charge noise. Denoting the charge dispersion\cite{koch_charge-insensitive_2007} of level $j$ by
$\epsilon_j=\omega_j(n_g=0)-\omega_j(n_g=1/2)$, we therefore take the relative dephasing strengths to be
$\beta_j=2\epsilon_j/(\epsilon_1-\epsilon_0$). (The normalisation of $\alpha_j$ and $\beta_j$ is chosen to allow
the usual interpretation of $\gamma_1$ and $\gamma_\varphi$ as relaxation and dephasing rates for the first two
levels of the transmon spectrum.) In fact, the pure dephasing rate is sufficiently small for our qubits that we
set $\gamma_\varphi=0$. A comparison of additional simulations with auxiliary experimental results at increased temperatures allows us to place an approximate upper bound of $0.003$ on the number of thermal photons in the cavity.

For the steady-state solution of Eq.\ \eqref{master2}, the Hilbert space is truncated to a subspace with maximum
number of excitations $N$, using the projector $P_N=\sum_{0\le n+j\le N}\ket{n,j}\bra{n,j}$. In our simulations,
we keep up to $N=7$ excitations. To reach agreement with the experimentally measured signal for the strongest
drive powers, it is necessary to account for a small amount ($\sim-55\,\text{dB}$) of leakage of the drive past
the cavity. In addition, there is a small bias introduced by measuring the transmission as the square of the $I$
and $Q$ quadratures, each of which is subject to noise. Accordingly, the quantity that corresponds to the
experimental signal is $A^2=\lvert 2 \sqrt{\kappa}\tr(\rho_s a) + b \xi \rvert^2 + 2 \sigma_n^2$, where $b$
describes the leakage of the drive bypassing the cavity, and $\sigma_n$ is the measurement noise in each of the
$I$ and $Q$ channels.

Fits are obtained by minimizing the mean squared deviation between experiment and calculation over the full
power range, with fit parameters being $b$ and the two scaling factors describing the signal attenuation and
amplification for input and output signals. To obtain optimal agreement, we also make small adjustments to the
system parameters $\gamma_1$, $\kappa$, $g_0$, $\omega_\text{r}$, $\omega_{01}$, and $E_C$. These parameters are
confined by separate experiments to
narrow ranges, and all values used in fits are consistent within the experimental uncertainties. Once obtained, the same set of parameters was used in generating Figs.\ \ref{fig:powsweep}, \ref{fig:wolverine}, and the Supplementary Movies.\\

{\noindent\sffamily EXPERIMENT}\\
Measurements are performed in a dilution refrigerator at $15\,\text{mK}$. The sample consists of two
superconducting transmon qubits\cite{koch_charge-insensitive_2007,schreier_suppressing_2008}, coupled to an
on-chip coplanar waveguide (CPW) cavity. Fabrication of the sample followed the description given in ref.\
\onlinecite{schreier_suppressing_2008}. The  CPW resonator has a half-wavelength resonant frequency of
$\omega_\text{r}/2\pi = 6.92\,\text{GHz}$ and a photon decay rate of $\kappa/2\pi = 300\,\text{kHz}$.
Transmission measurements are performed using a heterodyne detection scheme. The transmitted RF voltage signal
through the cavity is mixed down to a 1\,MHz carrier signal, and then digitally mixed down to dc to obtain the
transmitted voltage amplitude as a function of frequency.  The vacuum Rabi coupling strengths for the two qubits
are obtained as $g_0/\pi =347\,\text{MHz}$ (qubit 1) and $g_0/\pi = 94.4\,\text{MHz}$ (qubit 2). Time domain
measurements of the qubits show that they are Purcell-limited  and completely homogenously broadened at their
flux sweet spots\cite{houck_controlling_2008}. The coherence times are $T_1 = 1.7\,\mu\text{s}$ and $T_2 =
0.7\,\mu\text{s}$ (qubit 1, away from the flux sweet spot) and  $T_1 = 1.4\,\mu$s and $T_2 = 2.8\,\mu$s (qubit
2, at flux sweet spot). The charging energies of the two qubits are measured to be $E_C/h = 400\,\text{MHz}$ and
$E_C/h=340\,\text{MHz}$. (See Supplementary Information, Methods, for further details.)

\bibliographystyle{nature}

\subsection*{Acknowledgments}
This work has been supported by Yale University via a Quantum Information and Mesoscopic Physics Fellowship (AAH, JK), the LPS/NSA-ARO grant  W911NF-05-1-0365, NSF grants DMR-0653377, DMR-0603369, and PHY-0653073, and by Academy of Finland. We thank J.\ Gambetta, A.\ Blais, and A.\ Wallraff for helpful discussions, and L.\ Frunzio and B.\ Johnson for fabrication of the sample.

\subsection*{Author contributions}
JMC led the experimental effort. LSB and JK performed the calculations and did most of the writing. AAH gave technical support and conceptual advice. ET contributed to the early theory. MHD, SMG, and RJS provided support and supervised the project.

\subsection*{Author information}
Correspondence and requests for materials should be addressed to RJS.

\end{document}